# An intense thermospheric jet on Titan

E. Lellouch[1], M.A. Gurwell[2], R. Moreno[1], S. Vinatier[1], D.F. Strobel[3], A. Moullet[4], B. Butler[5], L. Lara[6], T. Hidayat[7], E. Villard[8]

1. LESIA, Observatoire de Paris, Université PSL, CNRS, Sorbonne Université, Univ. Paris Diderot, Sorbonne Paris Cité, 5 place Jules Janssen, 92195 Meudon, France
2. Harvard-Smithsonian Center for Astrophysics, Cambridge, MA 02138, USA
3. Departments of Earth & Planetary Sciences and Physics & Astronomy, Johns Hopkins University, 3400 N. Charles Street, Baltimore, MD 21218, USA
4. SOFIA / USRA, NASA Ames Research Center, Moffett Field, CA 94035, USA
5. National Radio Astronomical Observatory, Socorro, NM 87801, USA
6. Instituto de Astrofisica de Andalucia – CSIC, Glorieta de la Astronomia, E-18008 Granada, Spain
7. Bandung Institute of Technology, Bandung, West Java, Indonesia
8. Joint ALMA Observatory, Alonso de Cordova, 3107 Vitacura, Santiago, Chile

**Winds in Titan's lower and middle atmosphere have been determined by a variety of techniques, including direct measurements from the Huygens Probe[1] over 0-150 km, Doppler shifts of molecular spectral lines in the optical, thermal infrared and mm ranges[2-4], probing altogether the ~100-450 km altitude range, and inferences from thermal field over 10 mbar - $10^{-3}$ mbar (i.e. ~100-500 km)[5-6] and from central flashes in stellar occultation curves[7-9]. These measurements predominantly indicated strong prograde winds, reaching maximum speeds of ~150-200 m/s in the upper stratosphere, with important latitudinal and seasonal variations. However, these observations provided incomplete atmospheric sounding; in particular, the wind regime in Titan's upper mesosphere and thermosphere (500-1200 km) has remained unconstrained so far. Here we report direct wind measurements based on Doppler shifts of six molecular species observed with ALMA. We show that unlike expectations, strong prograde winds extend up to the thermosphere, with the circulation progressively turning into an equatorial jet regime as altitude increases, reaching ~340 m/s at 1000 km. We suggest that these winds may represent the dynamical response of forcing by waves launched at upper stratospheric/mesospheric levels and/or magnetospheric-ionospheric interaction. We also demonstrate that the HNC distribution is restricted to Titan's thermosphere above ~870 km altitude.**

Observations of Titan were conducted on July 22 and August 18/19, 2016 with the Atacama Large Millimeter Array (ALMA). The period corresponded to Titan's Northern late Spring (solar longitude~ 80°). Observations altogether covered a fraction of the 329-364 GHz range, by means of three scheduling blocks, each associated with a frequency setup (see details in Supplementary Information). ALMA was at that time in a moderately extended configuration (C40-5), providing angular resolutions of ~0.16-0.23" ("natural weighting"), to be compared to 1.0"-1.05" for Titan and its ~1000 km thick atmosphere. The spectra, taken at spectral resolutions of 1-2 MHz, include the rotational signatures of nine molecular species (CO, HCN, HNC, $HC_3N$, $CH_3CN$, $C_2H_5CN$, $C_2H_3CN$, $CH_3CCH$ and $C_3H_8$, some of them with several isotopologues), mostly present in emission due to their being formed in Titan's stratosphere. As expected based on radiative-transfer model calculations, spectral signatures (see examples in **Supplementary Figure 2**) are best apparent at the limb, in relation with the gain in path and the decrease of the $N_2$ continuum. In addition, as already seen in previous ALMA and other datasets[4,10-11], most molecular emissions show non-uniform distributions across the disk. This results from the combined effects of latitudinal variations of temperature[5-6] and composition and of the observing geometry. In this manner our data bear information of the 3D temperature and composition fields, complementing the Cassini results[12-13] for some of the molecules (CO, HCN, $HC_3N$, $CH_3CCH$) and providing new information for the others.

Six molecules (HCN, DCN, HNC, $HC_3N$, $CH_3CN$, and $CH_3CCH$) were further observed at a high spectral resolution of ~60 kHz (~280 kHz for $CH_3CCH$) in order to map their precise line-of-sight (LOS) central frequencies. As exemplified on **Fig. 1** for HNC and $CH_3CN$, strong blue (red) shifts are evident at beam positions centered on Titan's east (west) equatorial limbs at 425 km altitude. These shifts correspond to a beam-integrated limb-to-limb speed difference of ~490 m/s for HNC and ~400 m/s for $CH_3CN$. This is much larger than Titan's solid rotation (16 m/s at 1000 km altitude) and implies very fast prograde winds. Maps of these Doppler winds are shown in **Fig. 2** for all six species (including separate fits for the components of the HCN hyperfine triplet), based on Gaussian fits of the line central frequencies at all beam positions with sufficient line signal (see details in Supplementary Information). For most molecules, wind information is restricted to regions at or beyond the Titan limb, as emission features are too weak in nadir geometry. A prominent exception is $CH_3CN$, for which the strong extended emission and the availability of three lines enable LOS wind measurements over the entire disk with an accuracy of ~6-8 m/s. For other molecules, LOS wind errors (at/beyond limb) are 6-8 m/s for $HC_3N$ and HCN, 12-20 m/s for HNC, and 30-50 m/s for the $CH_3CCH$ and DCN that exhibit weaker and broader lines.

Very strong prograde winds are seen in all species, with maximum measured LOS winds of 250-350 m/s, and even the less precise $CH_3CCH$ and DCN winds confirm this picture. **Fig. 2** reveals however differences in the wind regime from the different species. HNC LOS speeds are strongest at equator, forming a jet, with no hint for latitudinal asymmetry. In contrast, $CH_3CN$ winds are enhanced in the Southern (winter) hemisphere, as seen by the "widening" of the iso-wind contours when moving from South to North. None of the studied molecules shows convincing evidence for meridional or day-to-night flows, with upper limits estimated to be 25-30 m/s.

Assigning altitudes/pressure levels to these wind measurements relies on the so-called contribution functions, themselves dependent on the local vertical distributions of temperature and abundances and the pointing geometry. Ultimately, this would require analyzing our entire dataset in terms of the atmospheric thermal/composition 3D field. For now, we calculated typical contribution functions for equatorial nadir and limb conditions. Details on the procedure, involving inversion techniques for temperatures and abundances, are given in Methods and Supplementary Information. Note that for species whose lines are mostly Doppler-broadened (HNC and $HC_3N$), the classical inversion technique based on line profile was complemented by the limb sounding capabilities offered by the data. This novel approach provides us with the first determination of the HNC profile in Titan's atmosphere, found to be essentially restricted to thermospheric levels above ~870 km altitude. This has important implications for models[14]; in particular, the lack of HNC in the stratosphere suggests that Galactic Cosmic Rays play at most a minor role for Titan's HNC chemistry.

Equatorial profiles for all six species and the associated contribution functions are shown in **Fig. 3**. Together the various molecules probe the entire atmosphere above 150 km, encompassing the mesosphere (300-800 km) and thermosphere (> 800 km). The latter is best sounded by HNC which probes 900-1100 km, providing the first direct zonal wind measurement in that part of Titan's atmosphere. $CH_3CN$, DCN and $CH_3CCH$ sound the upper stratosphere/lower mesosphere at 150-400 km, while, in relation with their steep slope and large thermospheric abundance, HCN (main line at 354.505 GHz) and $HC_3N$ sound a broad region roughly centered at ~700 km and extending 200-300 km above and below, as detailed in Methods. The difference in the sounded altitudes is consistent with the similarities/differences in the structure of the winds for the different molecules. **Supplementary Figure 6** shows LOS winds as a function of projected distance from Titan's rotation axis. In spite of their larger uncertainties, winds inferred from DCN and $CH_3CCH$ have a similar structure as those from $CH_3CN$. In contrast, HNC data show a distinct signature of high speed winds at large projected distances from the polar axis, corresponding to the equatorial jet outlined above. Winds derived from $HC_3N$ and HCN bear similar structures, that appear intermediate between the previous ones (i.e. stronger winds than those from $CH_3CN$, but with a less prevalent jet compared to HNC). Simple fits of the LOS winds that assume solid-body rotation indicate that winds increase with altitude, from ~200 m/s equatorial at ~350 km to ~280-300 m/s at 600 km and above. The Richardson number associated with this wind profile remains comfortably above the Ri < ¼ limit for turbulence.

Assuming purely zonal flow, we used our forward model and a trial-and-error approach to de-convolve the LOS winds measured in $CH_3CN$, HNC and $HC_3N$ in terms of latitudinal wind profiles at the associated altitudes (see Supplementary Information). Results for $CH_3CN$, which pertain to $345^{+60}_{-140}$ km (0.01-0.7 mbar with peak sensitivity at 0.035 mbar) indicate a maximum 220 m/s speed at 15°S with, as expected from qualitative data inspection, a much more gradual decline in the Southern vs. Northern hemisphere. This result agrees within error bars with pioneering measurements[4] using the same species (160±60 m/s equatorial at 300±100 km, assuming solid-body rotation). The north/south asymmetry that is evident in our $CH_3CN$ data also confirms the well-trusted but still indirect inferences based on temperature fields and the use of the thermal wind equation. Those consistently indicate a strong, broad mid-latitude (~20°-50°) jet in the winter hemisphere with peak velocities in excess of 190 m/s, extending over ~1 decade in pressure centered at 0.1 mbar, while significant winds (up to 100 m/s) are still present at mid-latitudes of the opposite hemisphere[5-6]. Note that the thermal wind approach is valid only outside an altitude-dependent (but typically ± [10°-30°]) equatorial band, a limitation that does not hold in direct wind measurements. Analyses of stellar occultation central flashes in 2001-2003 (Northern winter solstice) also indicate maximum winds of ~200 m/s at 250 km, but disagree on their latitudinal extent[7-9].

For $HC_3N$ and HNC, the de-convolved equatorial speeds are 310-345 m/s, rapidly decreasing with latitude in both hemispheres and with no clear latitudinal asymmetry. For $HC_3N$, results are sharply different from Ref. [4], both in the estimate in the probed altitude ($710^{+190}_{-260}$ km here vs 450±100 km in [4]) and in the wind velocities (310 m/s equatorial here, vs 60±20 m/s in [4]). As the data from [4] refer to 2003-2004 (solar longitude L =268°-280°), the contradiction – if taken at face value – would imply a temporal variation of the mesospheric circulation, with winds decreasing (resp.

increasing) with altitude at southern (resp. northern) summer solstice, or near perihelion (resp. aphelion).

Based on INMS and CAPS/IBS data, ion winds at 1000-1400 km in the range 30-260 m/s (vector component along the spacecraft trajectory) were reported[15]. Above 1200 km, ions and neutrals are collisionally-coupled[16], so these speeds represent neutral winds, but the measured range was broad and the other wind components unconstrained. Our HNC data reveal that the circulation regime over 900-1100 km is characterized by a relatively broad (±30-35° latitude extent at half maximum) equatorial jet with ~350 m/s maximum speed, i.e. ~1.4 times the speed of sound. Such a regime was previously unexpected. Pre-Cassini models[17-18] predicted subsolar to anti-solar (SSAS) winds of ~50 m/s. Early Cassini/INMS density profiles were used[19] to construct an empirical altitude-latitude temperature model; solving the momentum equations, a circulation pattern with a zonal jet of ~50 m/s and very strong meridional winds (~150 m/s) at 70°N latitude was inferred. A more complete analysis of 32 Cassini/INMS passes[20] indicated that while mean temperatures over 1050-1500 km vary over 112-175 K, there is a conspicuous lack of correlation between temperatures, latitude, longitude, local time, and solar zenith angle, indicating that the thermospheric thermal structure and dynamics are not primarily controlled by the absorption of solar EUV. Some correlations were found between atmospheric temperatures and the local plasma environment[21,20], but appear to be of ambiguous interpretation. An analysis of thermospheric energetics[22] demonstrated that particle sources (magnetospheric ions, pick-up ions, and auroral electron precipitation) and Joule heating contribute lower power than EUV input – typically by a factor 10 below 1100 km, and solar EUV also dominates ionospheric production[23]. But Saturn's co-rotation electric field that drives magnetospheric plasma convection past Titan at 120 km/s also likely drives ~1 km/s ion convection in the ionosphere[24]. Collisionally-coupled ion/neutral winds of ~350 m/s would appear sufficient to power perpendicular electric fields $E_\perp = -v \times B$ of order 1.5 μV/m, as inferred[25].

Vertically propagating waves seem capable of depositing energy fluxes comparable to solar UV [Ref. 22]. Evidence for waves in Titan's mesosphere and thermosphere is abundantly present in stellar occultation profiles[9], in the Huygens/HASI profile[26] above 500 km, and in the INMS density/temperature profiles[22,27], and maybe in vertical molecular profiles. Gravitational tides[28] are not promising for explaining the strong wind speeds we measure, due to their small phase velocities (0 and ± 5.9 m/s equatorial) and inability to penetrate regions where their phase speed equals the zonal velocity (i.e. between 0-5 km and the 70-80 km region of minimum wind[1]). Above 300 km altitude where the radiative time constant becomes shorter than a Titan day[29] (= 16 Earth days), other gravity waves, in particular thermal tides generated by solar heating, diurnal and especially semi-diurnal with large vertical wavelength[30,31], will propagate vertically. They can deposit eddy momentum fluxes by viscous dissipation, interact with the weak solar-induced SSAS thermospheric circulation and produce a strong equatorial flow. Solar forcing of tides will be complex as a solar day for a parcel advecting with zonal winds is 16 days at the surface, approximately one day at 350 km and ~ 3/4 day at 1000 km. Detailed models are needed to assess the plausibility of the mechanism, but an analogy may hold with Venus' upper atmosphere, where gravity waves are thought to be responsible for the variable retrograde super-rotating zonal flow that superimposes to the strong SSAS regime[32].

**Corresponding author:** Emmanuel Lellouch. LESIA, Observatoire de Paris, Université PSL, CNRS, Sorbonne Université, Univ. Paris Diderot, Sorbonne Paris Cité, 5 place Jules Janssen, 92195 Meudon, France. emmanuel.lellouch@obspm.fr; 33-1-45077672.

**Acknowledgements** E.L., R.M., and SV acknowledge support from the Programme National de Planétologie (PNP-INSU). We thank Agustín Sánchez-Lavega and an anonymous reviewer for constructive comments. The paper is dedicated to the memory of Daniel Lellouch, deceased on February 13, 2019.

**Author Contributions** M.G. wrote the ALMA proposal. M.G. and R.M. reduced the ALMA data. E.L. analyzed and modelled the data (with contribution from R.M.) and wrote most of the manuscript. S.V. provided initial thermal profiles as well as insight in the general science context. D.F.S. led the interpretative part. All authors discussed the manuscript.

**Materials requests & Correspondence** should be addressed to the corresponding author.

**Competing interests** The authors declare no competing interests.

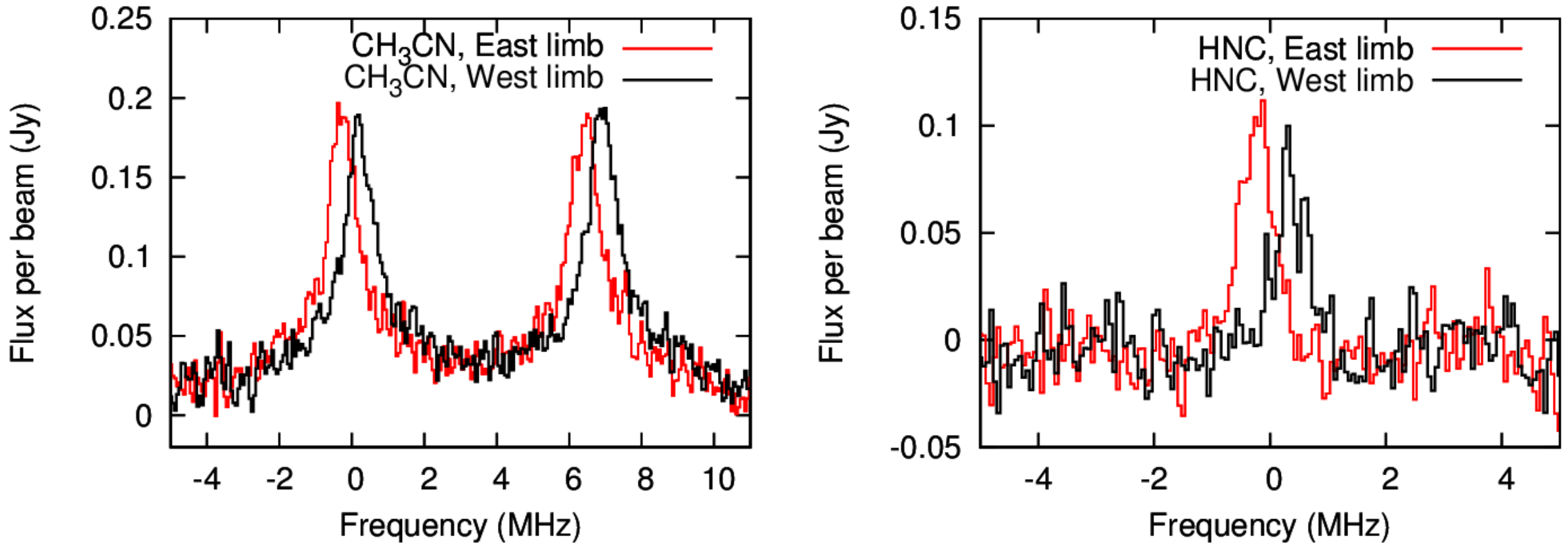

**Fig. 1**. **Evidence for zonal winds**. $CH_3CN$ (left) and HNC (right) line profiles at Titan East (+0.425", 0" from disk center) and West (-0.425",0" from disk center) limbs, observed on August 18, 2016. These pointings correspond to a radial distance of ±3000 km from disk center (i.e. a limb tangent altitude of 425 km above the surface). The reference zero frequencies are the rest frequencies of the $CH_3CN$ J, K=18,1 line (349.44699 GHz) and HNC (4-3) line (362.63030 GHz) respectively. The left panel also includes the $CH_3CN$ J, K=18,0 line at 349.45370 GHz. The limb-to-limb velocity difference is ~400 m/s for $CH_3CN$ and ~490 m/s for HNC.

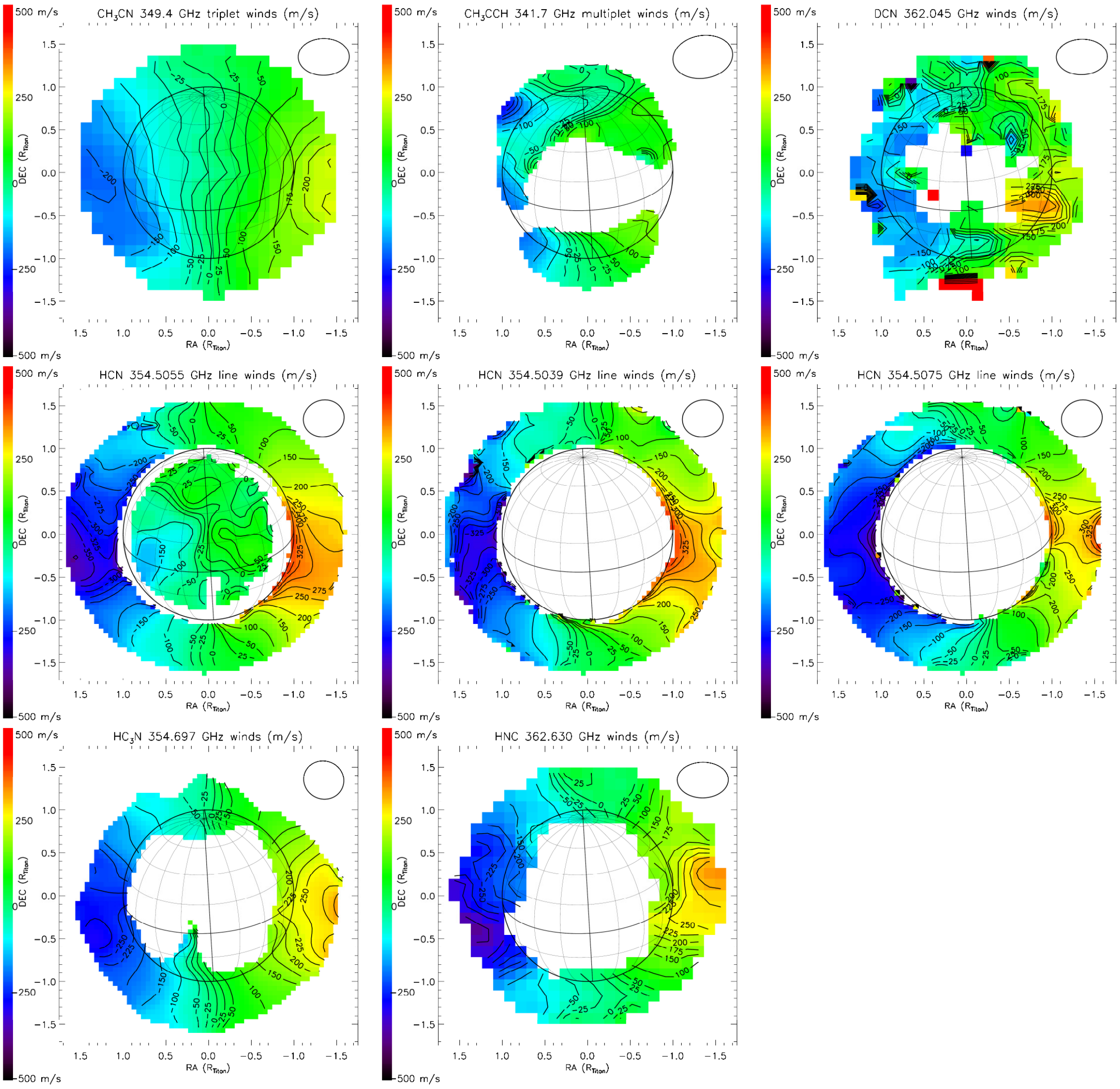

**Fig. 2. Doppler wind maps.** Line-of-sight Doppler shifts (m/s) measured in $CH_3CN$, $CH_3CCH$, DCN, HCN (main line of the hyperfine triplet at 354.5055 GHz and satellite lines at 354.5039 and 354.5075 GHz), $HC_3N$, and HNC. Right Ascensions (RA) and declinations are shown in terms of Titan's apparent radius. Ellipses show the synthetized beams. White areas correspond to regions where emission signals are too faint for reliable measurements to be performed. Only $CH_3CN$ emission makes it possible to measure winds throughout the disk. Some measurements within the disk are also possible using the main HCN 354.5055 GHz line, that appears in absorption in nadir geometry (see Supplementary Figure 2). Doppler shifts in the various species clearly indicate a prograde wind regime, but with different structure for $CH_3CN$ vs HNC, HCN and $HC_3N$. Winds in $CH_3CCH$ are mostly measurable outside of the low-latitude regions. Winds are only marginally detected from DCN emission, with 30-50 m/s error bars. See Supplementary Information for details.

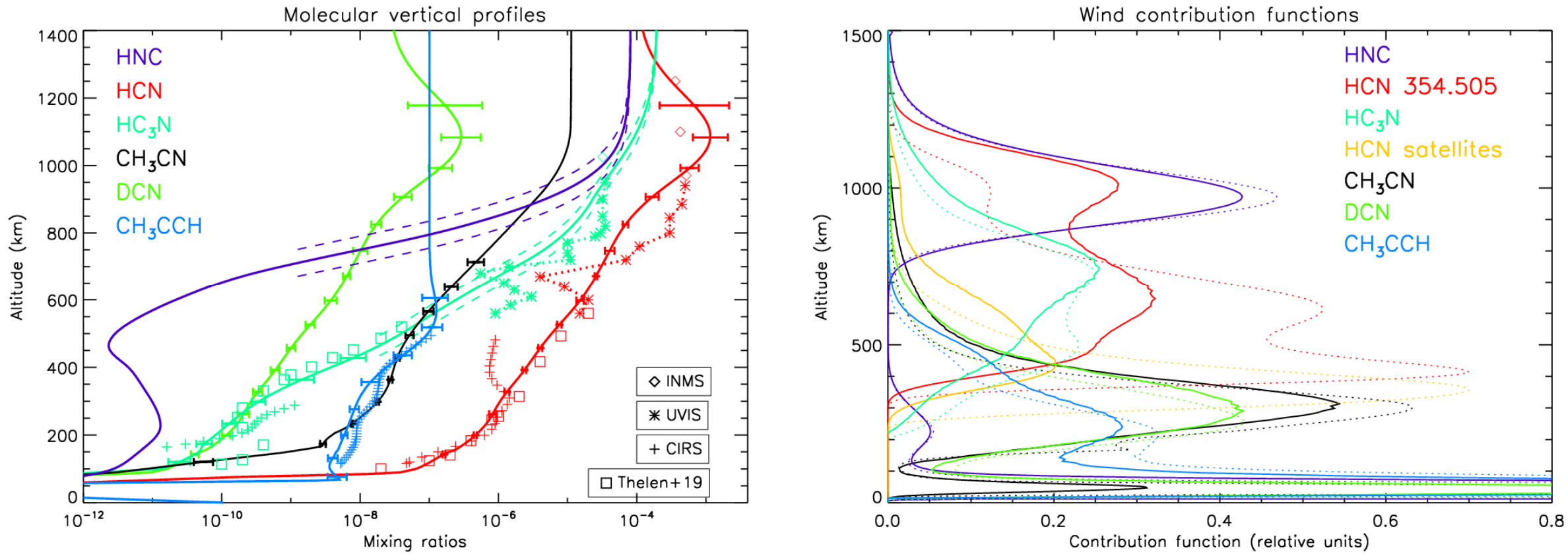

**Fig. 3. Molecular abundance profiles and probed altitudes. Left:** Equatorial vertical profiles for the six species used for the wind measurements (solid lines, with error bars indicated in the altitude ranges where information is available). As explained in Supplementary Information, uncertainties on HNC and on the upper part of the $HC_3N$ distribution are best described by a vertical shift of their altitude profile (dashed lines). For HCN, $HC_3N$ and $CH_3CCH$, characteristic profiles from Cassini/CIRS, UVIS and INMS, and from previous 2012-2015 ALMA observations (Thelen et al. 2019, Ref. [33]) are indicated. See Methods and Supplementary Information for details. **Right:** Wind contribution functions (WCF) in limb (solid lines) and nadir (dashed lines) geometries. For HCN, WCF are shown separately for the main line at 354.5055 GHz and for the satellite lines at 354.5039 and 354.5075 GHz (averaging the two). For $HC_3N$, to avoid the contribution from the nearby HCN (4-3) line, a WCF calculated with zero $HC_3N$ has been subtracted from the WCF before plotting. For other species, the peak below 100 km is due to the continuum. See Methods for details.

## Methods

### 1) Radiative transfer models

Spectral calculations were based on a home-made radiative transfer model in full spherical geometry, inherited from [4, 34, 35]. Line opacities are calculated using spectroscopic parameters from the Cologne Database for Molecular Spectroscopy (CDMS, https://www.astro.uni-koeln.de/cdms). As a consequence of small particle size and sharply decreasing absorption coefficients beyond 100 μm[36], thermal emission and scattering by haze are neglected, as in all previous studies of Titan at mm/submm wavelengths. Pencil-beam radiances are calculated on a grid of impact parameters (distance to Titan center) ranging from 0 to 4120 km (1.6 Titan radius). For each beam position under study, line-of-sight radiances as a function of impact parameter from Titan's center are re-interpolated over the Titan + atmosphere disk, and integrated over portions of the disk intercepted by the ALMA synthesized beam. At this step, we took into account the beam's elliptical shape and its orientation, and the precise apparent radius of Titan for each individual observation. This resulted in synthetic spectra expressed in Jy/beam which could then be directly compared to the data, after applying whenever needed minor (< 10 %) model-based calibration corrections to the data fluxes.

### 2) Temperature and abundance retrievals

Temperature profile information was derived from modelling of the CO (3-2) and HCN (4-3) lines and their isotopic variants, using a home-made inversion code[35] and based on optimal estimation methods principles[37-38]. For the purpose of this task, which is mainly to assign altitude levels to the wind measurements, we focused on the equatorial region, leaving the study of the thermal and compositional field for future work. For the a priori, we used an equatorial profile derived from Cassini/CIRS measurements in May 2016 (T119 flyby), extended above 700 km at a 153 K temperature, typical of the upper atmosphere according to Cassini / INMS measurements, and shown in **Supplementary Figure 1** (black line). While CO lines constrain atmospheric temperatures below ~350 km, HCN probes much higher altitudes (up to ~1000 km), with the difficulty that the temperature profile *and* the HCN profile are poorly constrained above the top of the sensitivity range of CIRS (~520 km). To alleviate the ambiguity between abundance and temperature effects in the core of the HCN (4-3) line, we inverted simultaneously the equatorial HCN (4-3) and $H^{13}CN$ (4-3) lines near the central meridian *and* at the limb. Indeed, the characteristic absorption / emission appearance of the HCN hyperfine structure in nadir / limb geometry provides complementary information on temperature and abundance sensitivity. In doing so, we used the HCN profile from Ref. [39] (extended above 400 km to reach HCN = $10^{-4}$ at 800 km and above) as a priori.

The best fit equatorial thermal profile retrieved in this manner is shown in **Supplementary Figure 1** (blue line), and the associated fits of the CO, HCN and $H^{13}CN$ lines in nadir and limb geometry are shown in **Supplementary Figure 2**. The temperature profile shows a marked mesopause at 140 K near 800 km. The occurrence of such a cold temperature somewhere above the stratopause is in fact directly implied by the depth of the main component of the HCN (4-3) hyperfine structure, with a brightness temperature of 143 K in the line core in nadir geometry. Note also that the best fit temperature / HCN profile permits a good match of the HCN v2=1 emission at 354.460 GHz (**Supplementary Figures 2 and 3**). The retrieved profile is also warmer than the a priori (CIRS) profile by up to ~3 K at 250 km and colder by ~6 K at 400-600 km. Differences between CIRS- and ALMA CO-derived thermal profiles of up to 5 K below 300 K have already been noted[40].

Error bars on the retrieved HCN/temperature retrievals are a combination of (i) uncertainties due to measurement noise and (ii) uncertainties due to the a priori used in the inversion process. Noise-related uncertainties were calculated using the formalism described in Ref. [37] (their eq. 23 and 24). The associated temperature errors are formally very small (0.2-0.3 K) in relation to the high S/N and high spectral-resolution of the data. Uncertainties due to the a priori were evaluated by repeating the temperature/HCN inversion process using several a priori temperature and HCN profiles, as detailed in Supplementary Information. These modelling variants yielded a suite of solution profiles, whose dispersion is a good indicator on the actual error bars on the retrieved temperature and HCN profile **(Supplementary Figure 4).** Beyond the well-known characteristics of Titan's middle atmosphere thermal structure such as the stratopause near 270 km, robust new features of all retrieved profiles are (i) the existence of a quasi-isothermal region over 450-600 km (already apparent in the CIRS-based a priori) and (ii) the existence of a ~140 K mesopause, located at 800±50 km altitude depending on the a priori. Error bars on the retrieved temperatures remain below 2-3 K up to 800 km; above this level they increase with the progressive lack of sensitivity of the measurements to ultimately reflect the a priori errors. The HCN profile (**Fig. 3**) is also very well defined (to within ~10-20 %) over 100-850 km, before error bars progressively increase. The nominal HCN mixing profile we determine reaches a $6.3 \times 10^{-4}$ mixing ratio at 1000 km, in agreement with results from both Cassini/INMS[41] and Cassini/UVIS[42], respectively above and below 950 km (see Fig. 3).

The retrieved thermal profile shows a quasi-isothermal part at 156 K over 450-600 km and a 140 K minimum near 800 km. Although the resulting oscillatory structure may be reminiscent of the temperature perturbations seen in the Huygens/HASI data[26], its attribution to a wave is doubtful because (1) our data sample broad regions (beam size ~1000 km), and nadir and limb data inverted simultaneously are distant by ~2500 km (2) the resulting wave amplitude and vertical wavelength would be ~5-10 K and ~300 km, respectively smaller and larger than in the HASI data (~10-20 K and ~100 km). Instead, the temperature profile, and in particular the 800 km temperature minimum, may be mostly radiatively-controlled. A pioneering non-LTE model[43] of Titan's upper atmosphere predicted a 135-140 K mesopause (but at 600 km), primarily due to $C_2H_6$ cooling. Revising such a model in the light of new distributions derived from Cassini and the present work (especially for HCN, $C_2H_6$, $HC_3N$ and $CH_3CN$) and with updated spectroscopic databases would shed further light on this issue.

The retrieved thermal profile was then used to derive the other molecular profiles, again for the equatorial region. The DCN profile was taken as the retrieved HCN profile, multiplied by a constant DCN/HCN ratio of $(2.5\pm0.2) \times 10^{-4}$ [44], i.e. assuming that the DCN profile follows that of HCN at all altitudes. $CH_3CN$ and $CH_3CCH$ were inverted using a priori profiles respectively based on ground-based data[39] (extended to $10^{-5}$ at 1000 km and above) and CIRS observations[45]. For these three species, best fits of the (west) limb equatorial emission are presented in **Supplementary Figure 3**. Error bars on $CH_3CN$ and $CH_3CCH$ reflect the combined effects of noise, temperature profile uncertainty and a priori errors.

The Doppler-limited line-shape of HNC J=4-3 line (**Supplementary Figure 5**) contains limited information on the HNC vertical profile, except for the fact that this precludes HNC to be present at too deep atmospheric levels. Based on HNC J=6-5, ref. [34] found that for constant mixing profiles above some level, the cutoff altitude cannot be lower than 300 km. Using HNC data averaged over a concentric ring 2000-3350 km from Titan center, and testing photochemical profiles[14] for HNC, we found that several of these HNC profiles overestimate the amount of HNC in the lower stratosphere. In contrast, the model in which the production of HNC by Galactic Cosmic Ray impact is omitted is essentially in agreement with the observed spectrum. While this line shape-based approach does not constrain the distribution of HNC at altitudes above the pressure-broadening levels, the latter can be determined from the observed variation of the line-integrated intensities with radial distance. Comparing the latter with model calculations for constant HNC profiles above some cutoff altitudes, we found that the bulk of HNC is located above 870±40 km altitude, implying that HNC is mostly a thermospheric species. The photochemical model from Ref. [14] without GCR also provides the correct radial dependence for the HNC emission, and was adopted as nominal profile, allowing for a vertical shift of ±40 km in its altitude scale.

The situation is similar for the $HC_3N$ J=39-38 emission at 354.697 GHz, which – except in the high-northern and southern latitude regions where it is strongly enhanced – does not show Lorentzian wings. We performed the same study as for HNC on the radial distance variation of the $HC_3N$ emission (using the slightly stronger J=38-37 line at 345.609 GHz observed along with CO (3-2)), restricting ourselves to low-latitude regions. Considering piecewise profiles, this approach yielded a best fit cutoff altitude of 630±30 km (**Supplementary Figure 5**), implying that $HC_3N$ too is mostly present in Titan's upper atmosphere, albeit less so than HNC. This is not surprising given that $HC_3N$, as measured by CIRS[12-13] has also a measurable abundance in Titan's stratosphere and lower mesosphere, with mixing ratios increasing from $6 \times 10^{-11}$ – $1 \times 10^{-7}$ over 150-450 km. Initializing the $HC_3N$ profile at these values and a large abundance at z > 800 km, we determined an equatorial $HC_3N$ profile fitting both the shape of the J=39-38 line and the radial variation of the J=38-37 lines fluxes. In contrast, the nominal $HC_3N$ profile from Ref. [14] does not fit well any of either constraint, indicating that this profile is overabundant at lower altitudes and under-abundant in the upper atmosphere. Our $HC_3N$ solution profile has a 1000 km abundance as high as $\sim 5 \times 10^{-5}$. This agrees rather well with the global value reported from Cassini/INMS[46] and with the Cassini/UVIS $HC_3N$ abundance over 850-950 km [42]. We reiterate that our conclusions pertain to the equatorial $HC_3N$ profile only.

**Fig. 3** summarizes the retrieved profiles for all six molecules used to probe the winds. Error bars are indicated in the range where information is available. For HNC and $HC_3N$, uncertainties are best expressed as a vertical shift of ±40 km and ±30 km in the profiles, for reasons explained above. These profiles are compared to characteristic profiles locally derived from Cassini/UVIS at 6°S (T41I, Ref. [42]), INMS[41,46], and CIRS. For the latter, we show HCN and $CH_3CCH$ profiles at 9°S from May 2016 (T119, Ref. [45]); for $HC_3N$, not clearly detected in the latter observations, we use instead a 5°N profile from Jan. 2007 (T23, Ref. [47]). Also shown in Fig. 3 are disk-averaged HCN and $HC_3N$ profiles derived from 2012- 2015 ALMA observations (Ref. [33], their Fig. 10). **Fig. 3** highlights the overall agreement between these profiles, giving in particular great confidence in our HCN and $HC_3N$ profiles in the middle/upper atmosphere. A noteworthy difference is that "local" molecular profiles (from Cassini CIRS and UVIS) show small-scale vertical variations (see e.g. Figs. 3- 4 of Ref. [47], and Fig. 26 of Ref. [42]), not predicted in photochemical models[14] and presumably of dynamical origin; those are particularly visible in **Fig. 3** for the HCN and $HC_3N$ profiles derived from UVIS.

### 3) Contribution functions and wind altitudes

The retrieved profiles for all six molecules used to probe the winds are shown in **Fig. 3** (left panel). Based on these profiles, "wind contribution functions" were calculated to estimate the range of altitudes probed by the wind measurements in each molecule. For this, we first calculated "traditional" contribution functions $B(T)\ \partial \exp(-\tau) / \partial \ln(p)$, both monochromatically (over a spectral interval matching the one used for the wind measurements), and for every pencil-beam line-of-sight. At each frequency, these contributions functions were then convolved with the synthetized beam associated to each observation. To estimate how the probed altitude may vary with the observing geometry, these spatially-convolved – but still monochromatic – contribution functions were calculated both in nadir view (i.e. pointing to Titan's disk center) and in limb view, defined here as a position distant from Titan's center by 1.3 Titan radius (i.e. 772.5 km above the solid planet limb). This corresponds typically to the distance where we measure strongest winds before signals get too faint. Finally, these spatially-convolved contribution functions were spectrally convolved, using a weight proportional to the local spectral slope of the synthetic spectrum[48,4]. For each species, these wind contribution functions (WCF) for the nadir and limb geometries are shown in the right panel of **Fig. 3**. For HCN, we show separately the WCF for the main line at 354.5055 GHz and for the two satellite lines at 354.5039 and 354.5075 GHz (averaging their two very similar WCF).

Wind contribution functions highlight three regions: (i) a narrow (900-1100 km) thermospheric region probed by HNC (ii) another relatively narrow region, encompassing the upper stratosphere / lower mesosphere (200-400 km) probed by $CH_3CN$, $CH_3CCH$, and DCN (iii) a much broader zone, straddling from ~400 to ~1000 km (in limb geometry), probed by HCN and $HC_3N$. In limb geometry, the mean probed altitudes (i.e. the WCF-weighted mean altitude) are: $345^{+60}_{-140}$ km for $CH_3CN$; $315^{+90}_{-160}$ km for $CH_3CCH$; $340^{+90}_{-155}$ km for DCN; $745^{+365}_{-320}$ km for the main HCN line; $540^{+120}_{-185}$ km for the HCN satellite lines; $710^{+190}_{-260}$ km for $HC_3N$; and $990^{+100}_{-120}$ km for HNC, where the "error bars" mark the vertical extent of each WCF, i.e. the range in which it exceeds ½ of its maximum value. The impact of the thermal profile and abundance uncertainties (see Supplementary Information) on the wind contribution function (WCF) are small. Specifically, the mean probed altitude, as defined above, shifts by ±45 km for HNC, ±20 km for $HC_3N$, ±25 km (resp. ±15 km) for the main (resp. the satellite) HCN line, and ±10 km for $CH_3CN$, $CH_3CCH$ and DCN. Thus, except to some degree for HNC, the associated error bar appears dwarfed by the broad vertical extent of the WCF, from which we conclude that the vertical scale associated with our wind measurements is robust.

Wind contributions functions calculated by [4] for Plateau de Bure interferometer observations of $CH_3CN$ and $HC_3N$ lines at 220 GHz indicated sounding altitudes of 300±150 km for $CH_3CN$ and 450±100 km for $HC_3N$. This compares well with our computation for $CH_3CN$ but is rather different for $HC_3N$. This stems from the combination of three factors, all of which "pushing" the contribution function to higher levels in our case: (i) a higher $HC_3N$ abundance in the upper atmosphere compared to the profile[39] adopted in Ref. [4] (ii) the fact that we calculated the WCF for pointing towards a distance of 1.3 Titan radius (vs 1.0 Titan radius in [4]), and (ii) the ~3 times smaller synthesized beam in our observations (~0.2" vs 0.6"), which enhances the contribution of limb with respect to nadir, thereby tending to probe higher altitudes.

**Data availability** The data that support the plots within this paper and other findings of this study are available from the ALMA archive (http://almascience.nrao.edu/aq/) and from the corresponding author upon reasonable request.

**Code availability** Novel methods developed for this research and not available from previous studies are available from the corresponding author upon reasonable request.

## Additional references

# Supplementary Information for:

# An intense thermospheric jet on Titan

E. Lellouch, M.A. Gurwell, R. Moreno, S. Vinatier, D.F. Strobel,
A. Moullet, B. Butler, L. Lara, T. Hidayat, E. Villard

correspondence to: emmanuel.lellouch@obspm.fr

## 1) Observations and data reduction

We observed Titan as the focus of Cycle 3 program 2015.1.01023.S (PI. M. Gurwell). Titan was observed in three scheduling blocks (SB) named (A-C), one of them being scheduled on July 22, 2016 and the other two consecutively on August 18/19, 2016. The quasars J1634-2058 and J1626-2951 were used for phase calibration and flux calibration. All three scheduling blocks made use of Band 7 receiver sets, altogether covering large portions of the 329-364 GHz range. We employed a range of spectral resolutions. Typically, moderate 1-2 MHz resolutions were used to capture profiles of broad lines such as CO and HCN, while being also well suited to weak emissions. Higher resolutions (61 – 282 kHz) were also used, to study narrow yet strong lines, in particular, HNC, $HC_3N$, and the HCN hyperfine structure core, and with the intent of measuring winds from Doppler shifts. The on-source time was 49 min for each of the scheduling blocks. 38 antennas were active in July 2016 and 39 in August. Observing conditions were good ($H_2O$ = 0.44-0.66 mm) resulting in rms noise of 2.0 mJy/beam/MHz. Observing details are given in Supplementary Table 1.

Data reduction was initially carried out under CASA. Once calibrated visibilities were produced, deconvolution and imaging were performed under both the CASA and GILDAS environments. Self-calibration was performed to improve imaging quality, and the data were finally re-gridded on 0.02" or 0.05" pixel grids, expressing the spectral axis in Titan's rest frame. The achieved angular resolution ("natural weighting") varies from almost circular ~0.17" to more elongated 0.29" x 0.20" beams, as detailed in Supplementary Table 1. Titan's 5150-km diameter subtended 0.7594" and 0.7283" in the sky for the July and August observations respectively, and coordinate scales for maps were expressed in units of Titan's radius. For both periods, the sub-earth latitude was 26.0° and the North polar angle 3.45°.

## Supplementary Table 1: Observational summary

| SB | Date (UT) at start | Δ (au)[1] | Sub-earth long. | $T_{int}$ (min) | Synthesized beam (major, minor, orientation[2]) | Spectral range (GHz) | Spectral resolution (MHz) |
|---|---|---|---|---|---|---|---|
| C | July 22 2:23 | 9.3508 | 319° | 49 | 0.29"x 0.20" PA=-85° | 329.063-330.937 | 1.13 |
| | | | | | 0.29"x 0.20" PA=-86° | 330.826-332.700 | 1.13 |
| | | | | | 0.26"x 0.19" PA=-81° | **341.448-341.917 ($CH_3CCH$ J=19-18)** | **0.282** |
| | | | | | 0.28"x 0.19" PA=-81° | 342.263-343.196 | 0.98 |
| B | August 18 23:48 | 9.7504 | 230° | 49 | 0.23"x0.16", PA=91° | 349.161-349.626 | 0.25 |
| | | | | | 0.23"x0.16", PA=91° | **349.419-349.474 ($CH_3CN$ J=19-18)** | **0.0625** |
| | | | | | 0.22"x0.15", PA=90° | **362.018-362.073 (DCN J=4-3)** | **0.0625** |
| | | | | | 0.22"x0.15", PA=90° | **362.602-362.657 (HNC J=4-3)** | **0.0625** |
| | | | | | 0.22"x0.15", PA=90° | 361.948-363.815 | 1.0 |
| A | August 19 1:16 | 9.7512 | 230° | 49 | 0.18"x0.16", PA=80° | 342.363-344.136 | 1.13 |
| | | | | | 0.18"x0.16", PA=80° | 344.063-345.936 | 1.13 |
| | | | | | 0.17"x0.16", PA=74° | **354.477-354.534 (HCN J=4-3)** | **0.061** |
| | | | | | 0.17"x0.16", PA=75° | 354.260-356.125 | 1.95 |
| | | | | | 0.17"x0.16", PA=72° | **354.677- 354.737 ($HC_3N$ J=39-38)** | **0.061** |

[1] Geocentric distance [2] Polar angle of beam major axis

Entries in bold face indicate data used for wind measurements

## 2) Doppler wind measurements

Wind measurements were obtained from inspection of Doppler shifts on each individual spectra in the maps. For this, we performed Gaussian fits (the sum of a Gaussian and a 1$^{st}$ or 2$^{nd}$ degree polynomial) of the line profiles in their central parts, over ranges depending on the linewidth for each species. The fitting interval was ± [1.875 MHz] for HNC and $HC_3N$ and ± [3.75 MHz] for DCN. For HCN, which shows hyperfine structure, the spectrum was fit over ± [3.6 MHz] from the main component at 354.5055 GHz, by the sum of three Gaussians, in order to independently determine Doppler shifts for both the main component and for the two satellite lines at 354.5039 and 354.5075 GHz. For $CH_3CN$, we performed separate fits for the three lines at 349.427, 349.447 and 349.454 GHz, considering each time a ± [2.5 MHz] interval. Results were very similar for each line, and then averaged over the three lines. For $CH_3CCH$, due to lower line contrasts, we first created at each beam position a fictitious "total" line, by realigning the spectra successively on the five strongest lines of the 341.7 GHz multiplet (i.e., the 341.637, 341.682, 341.715, 341.735, and 341.741 GHz lines), and averaging them with weights proportional to the individual line strengths. The resulting "total" line was then fit for its position over a ± [8.5 MHz] interval. Wind measurements were obtained in this manner at all points of the map where the spectra showed sufficient emission signals for the Gaussian fits to be meaningful. Depending on the species, this was possible when the line signal exceeded 10 to 35 % of its maximum value on the map. As a result, wind measurements are usually reliable on a more or less extended region centered on the limb, an exception being $CH_3CN$ for which emission is also strongly detected at all nadir (disk-intercepting) beam positions, enabling wind maps on the entire Titan disk. For HCN, we also obtained some wind measurements within the disk using the main component of the hyperfine triplet, that appears in absorption against the disk (Supplementary Figure 2). To attribute error bars on the winds, we used a simple Monte-Carlo approach[48]. Residuals between the observed spectra and the Gaussian fits were used to estimate a rms noise level. Random noise at this level was then superimposed on the Gaussian model, which was then refit. Repeating this procedure 300 times, the dispersion in the fitted positions provided an estimate of the error bars on the LOS wind measurements. In the case of $CH_3CN$, the error bars were further decreased by averaging the results on the three components. Wind maps for the 6 species are shown in Fig. 2. Typical error bars for measurements at/beyond the limb are 12-20 m/s on HNC, 6-8 m/s on $HC_3N$, $CH_3CN$, and the HCN main line, 10-12 m/s for the two HCN satellite lines, and 30-50 m/s on DCN and $CH_3CCH$. Errors bars for LOS winds based on the HCN main line within the disk are ~20 m/s.

## 3) Temperature and abundance retrievals

Temperature profile information was derived from modelling of the CO (3-2) and HCN (4-3) lines and their isotopic variants, using a home-made inversion code[35] and based on optimal estimation methods principles[36-37]. For the purpose of this task, which is mainly to assign altitude levels to the wind measurements, we focused on the equatorial region, leaving the study of the thermal and compositional field for future work. For the a priori, we used an equatorial profile derived from Cassini/CIRS measurements in May 2016 (T119 flyby), extended above 700 km at a 153 K temperature, typical of the upper atmosphere according to Cassini / INMS measurements, and shown in Supplementary

Figure 1 (black line). We first used the $^{12}CO$ (3-2) and $^{13}CO$ (3-2) lines observed at Equator and near the central meridian to retrieve the CO mixing ratio (48±1 ppm) and an intermediate-step thermal profile. The latter (red line in Supplementary Figure 1) deviates from the a priori only below ~350 km, marking the top of the sensitivity layer to CO. HCN probes much higher altitudes, however (up to ~1000 km), but the difficulty is that the temperature profile *and* the HCN profile are poorly constrained above the top of the sensitivity range of CIRS (~520 km). Thus in a second step, keeping the temperature profile fixed below 350 km, we performed a simultaneous inversion of the HCN (4-3) and $H^{13}CN$ (4-3) lines in terms of temperature and HCN profiles. We assumed a telluric-like HCN / $H^{13}CN$ ratio of 89, consistent with other measurements of carbon isotopes in Titan's atmosphere, and took the HCN profile from [39] (extended above 400 km to reach HCN = $10^{-4}$ at 800 km and above) as a priori. When doing so, to alleviate the ambiguity between abundance and temperature effects in the core of the HCN (4-3) line, we inverted simultaneously the equatorial HCN (4-3) and $H^{13}CN$ (4-3) lines near the central meridian *and* at the limb. Indeed, the characteristic absorption / emission appearance of the HCN hyperfine structure in nadir / limb geometry provides complementary information on temperature and abundance sensitivity. In practice, we used the west limb spectrum centered at RA, DEC = (0", -0.42") from Titan center, i.e. 400 km above Titan's solid surface limb, correcting for the wind Doppler shift before performing the inversion.

The best fit equatorial thermal profile retrieved in this manner is shown in Supplementary Figure 1 (blue line), and the associated fits of the CO, HCN and $H^{13}CN$ lines in nadir and limb geometry are shown in Supplementary Figure 2. The temperature profile shows a marked mesopause at 140 K near 800 km. The occurrence of such a cold temperature somewhere above the stratopause is in fact *directly* implied by the depth of the main component of the HCN (4-3) hyperfine structure, with a brightness temperature of 143 K in the line core in nadir geometry. Note also that the best fit temperature / HCN profile permits a good match of the HCN v2=1 emission at 354.460 GHz (Supplementary Figures 2 and 3). The retrieved profile is also warmer than the a priori (CIRS) profile by up to ~3 K at 250 km and colder by ~6 K at 400-600 km. Understanding of these differences is deferred to future work, but we note for the time being that differences between CIRS- and ALMA CO-derived thermal profiles of up to 5 K below 300 K have been noted[40].

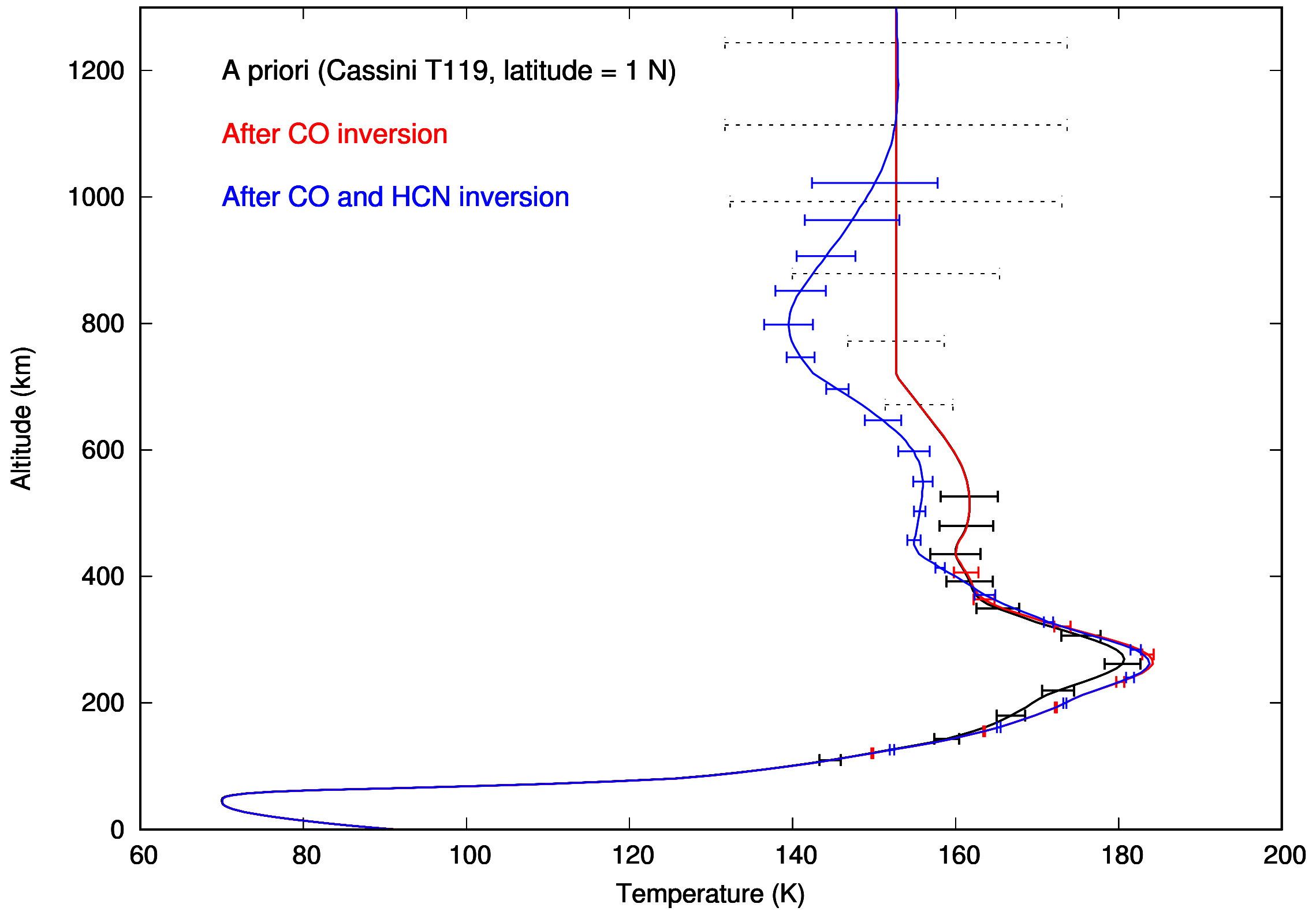

**Supplementary Figure 1. Thermal profiles**. Equatorial thermal profiles used in this study. Black: a priori profile, based on Cassini/CIRS measurements. Error bars indicate the adopted a priori errors (see text). They are shown in dotted lines above the region constrained by Cassini/CIRS. Red: temperature profile retrieved from inversion of $^{12}CO$ and $^{13}CO$. Blue: thermal profile retrieved from inversion of the HCN and $H^{13}CN$ limb and nadir data, using the red profile as a priori. Error bars on the retrieved profile include the propagation of the a priori errors on temperature and HCN (see text). This profile is then used to determine other molecular profiles and calculate contribution functions.

Error bars on the retrieved HCN/temperature retrievals are a combination of (i) uncertainties due to measurement noise and (ii) uncertainties due to the a priori used in the inversion process. Noise-related uncertainties were calculated using the formalism described in Ref. [37] (their eq. 23 and 24). The associated temperature errors are formally very small (0.2-0.3 K) in relation to the high S/N and high spectral-resolution of the data.

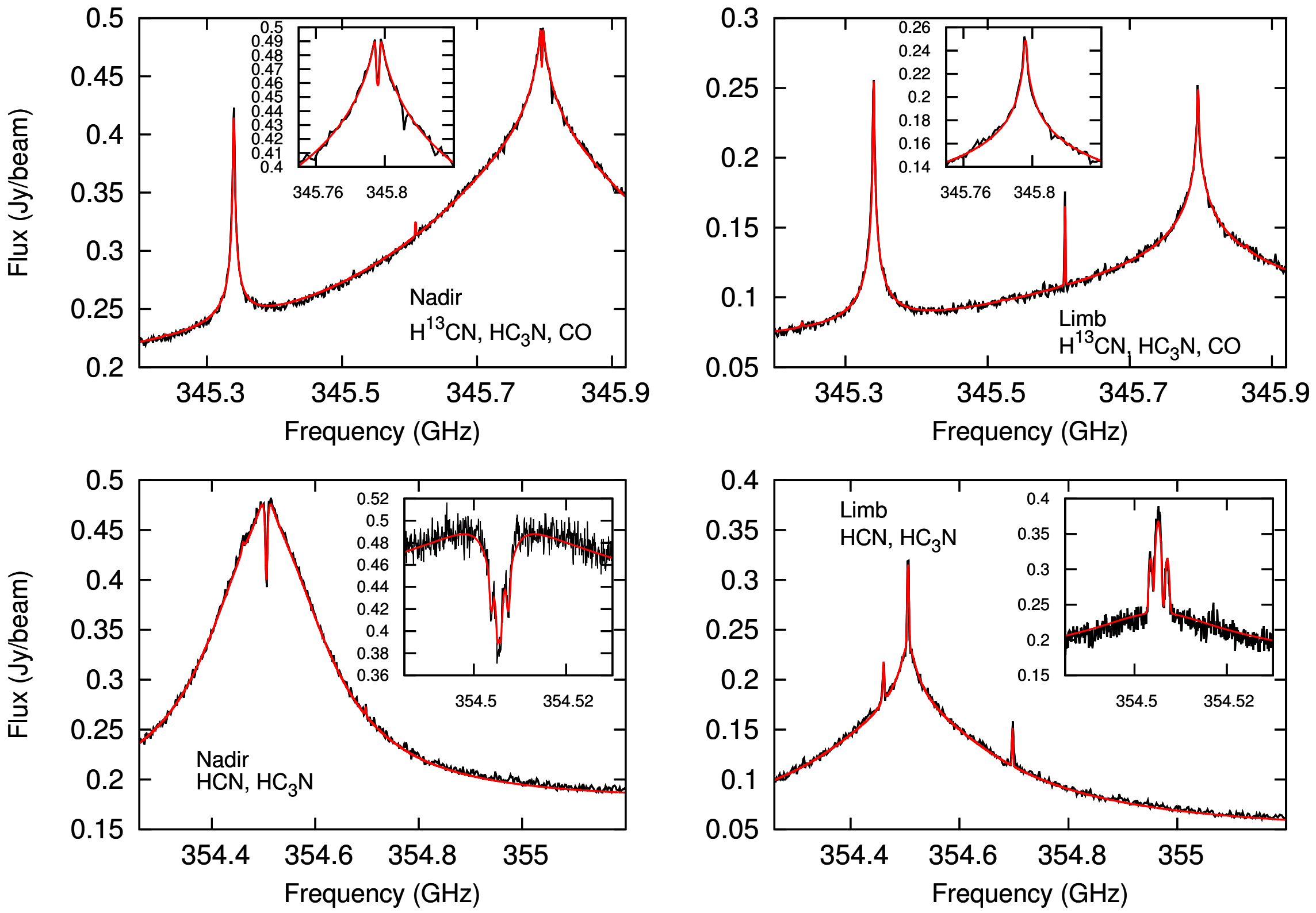

**Supplementary Figure 2. Spectral fits for CO and HCN**. Fits of the CO (3-2), HCN (4-3), and $H^{13}CN$ (4-3) lines at the equator and in nadir and limb geometries. The insets show the cores of CO (3-2) and HCN (4-3). Note also, mostly visible in limb geometry, the HCN $v_2$ line at 354.460 GHz and the $HC_3N$ J=38-37 and J=39-38 lines at 345.609 and 354.697 GHz. The associated retrieved profile for HCN is shown in Fig. 3.

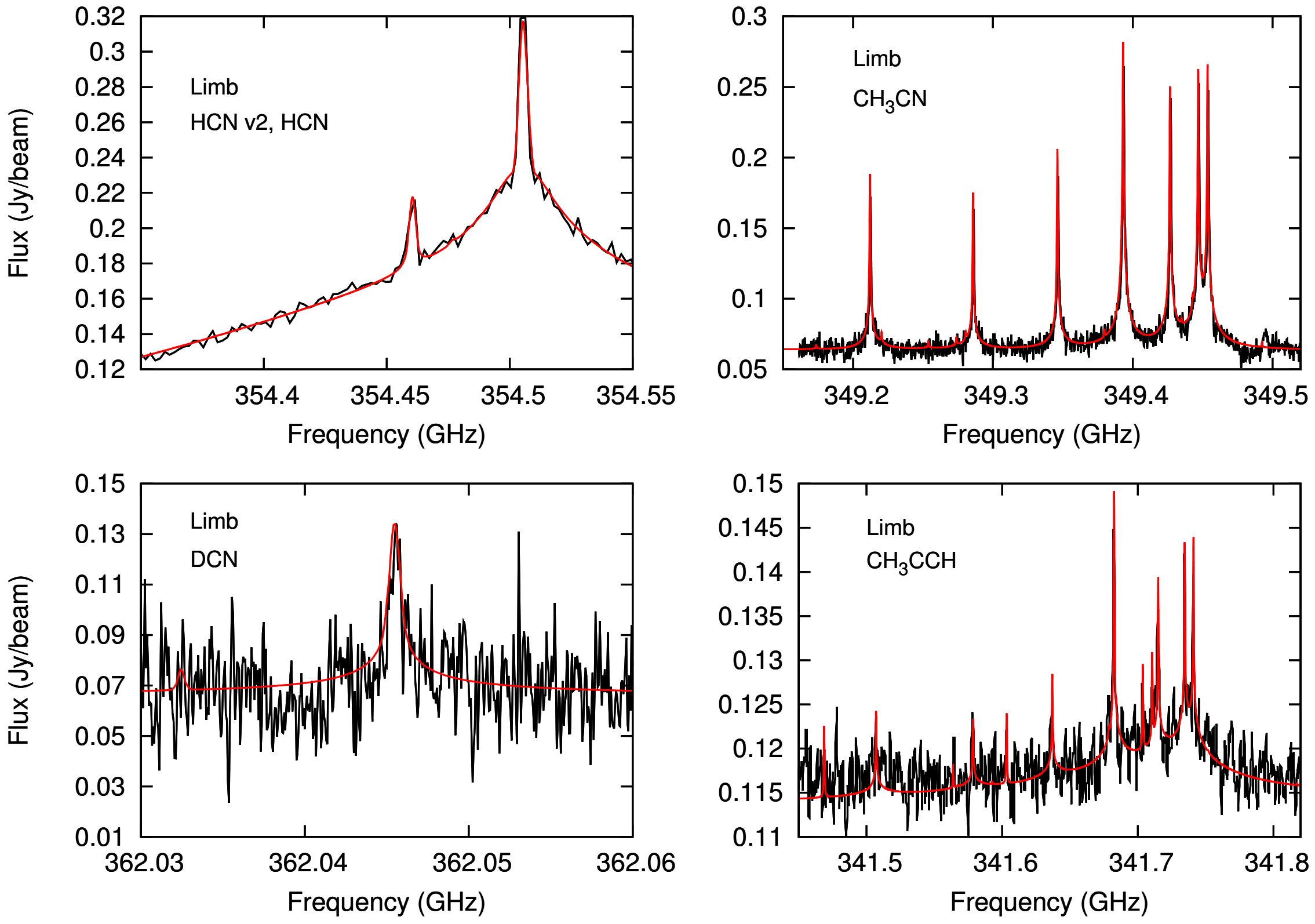

**Supplementary Figure 3. Spectral fits for other species**. Zoom of the HCN v2 line at 354.460 GHz and fits of $CH_3CN$, DCN, and $CH_3CCH$ equatorial limb data. The associated retrieved profiles are shown in Fig. 3.

To evaluate uncertainties due to the a priori, we repeated the temperature/HCN inversion process by using several a priori temperature and HCN profiles. For this, the first step was to construct a realistic range of initial temperatures profiles. As indicated above, we nominally used an equatorial profile derived from Cassini/CIRS measurements[45] in May 2016. The formal uncertainty on this profile is < 1 K over 7.5 mbar - 0.025 mbar (100-350 km), increasing to ~ 2 K over 5 x $10^{-3}$ – 1 x $10^{-3}$ mbar (430 - 520 km). As there are further systematic uncertainties on the CIRS-retrieved profile (e.g. related to the precise position of the CIRS detectors on the sky) at the level of ~1 K over 250-350 km, we conservatively adopted uncertainties increasing from 1 K at 100 km to 4 K at 630 km. Above this altitude, we adopted an a priori error that increases as log(pressure) to ~20 K at 1000 km and above. This choice was guided by (i) the observed dispersion in the INMS isothermal temperature measurements at 1050 -1500 km, which overall range from 112 K to 175 K around a mean temperature of 150 K over this altitude range (Ref. [20], Fig. 11), and by (ii) a similar dispersion in the few available temperature profiles from Cassini / UVIS (Ref.[42], Fig. 29). We note finally that temperatures at 1000 km from UVIS and HASI cluster around 180 K, vs 140 K for INMS; therefore, adopting a 153±20 K a priori temperature at 1000 km is reasonable. These a priori errors defined a set of three initial temperatures profiles for which the combined CO/HCN retrievals were performed (Supplementary Figure 1, black line with error bars).

We also studied the impact of using different a priori profiles for HCN. In practice, we used (a) HCN profiles enhanced or depleted from the nominal a priori[39] by constant factor-of-3 (b) an alternative a priori profile in which the HCN kept increasing with altitude up to $5x10^{-4}$ at 1000 km (matching the INMS[41] and UVIS[42] measurements near this altitude), allowing for a further factor-of-2 uncertainty in this profile. Altogether this provided six a priori HCN profiles which were also used (along with the nominal a priori temperature) for the inversion.

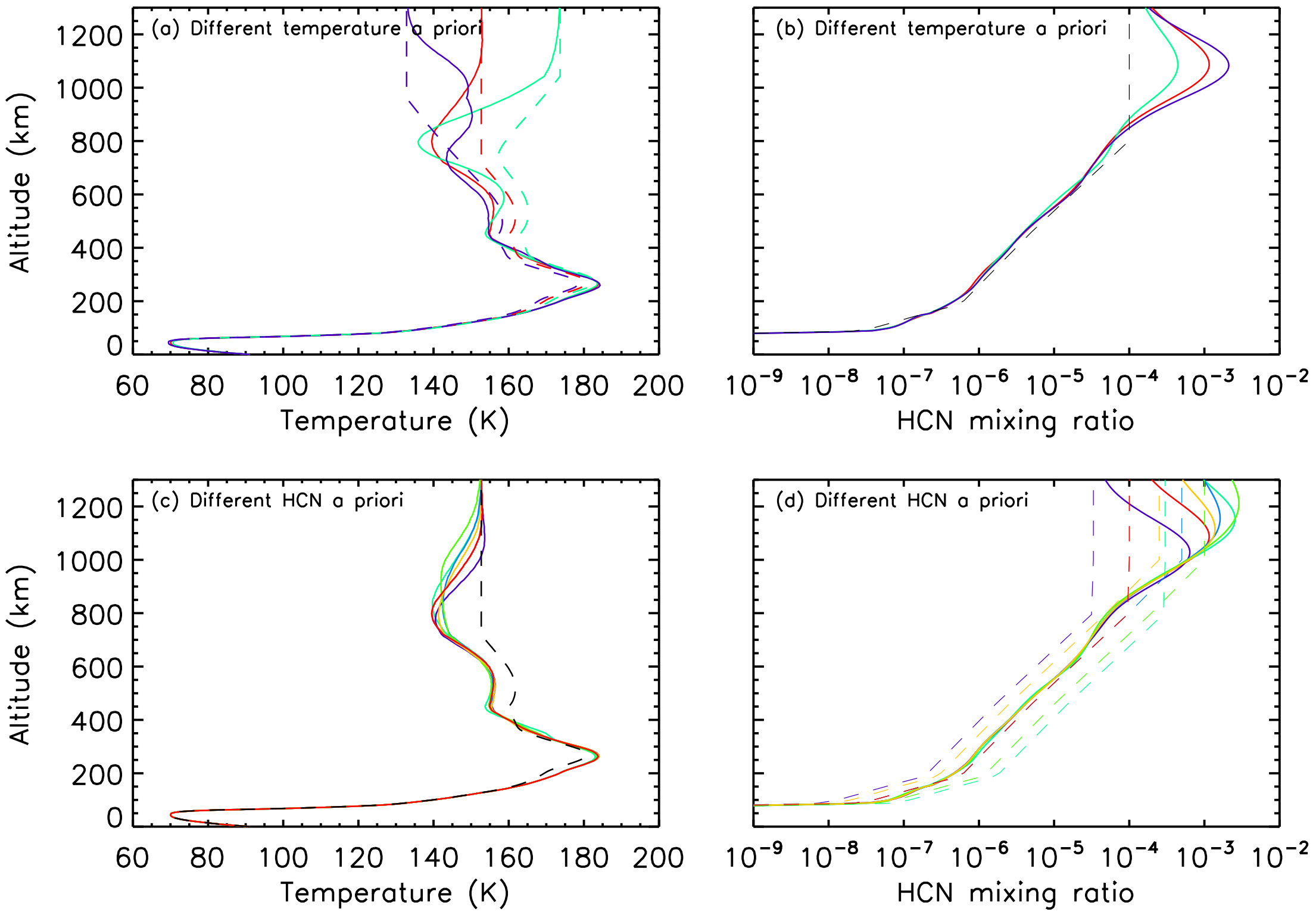

**Supplementary Figure 4. Sensitivity of retrievals to a priori temperature and HCN conditions.** Retrieved temperature and HCN profiles for a suite of a priori temperature profiles (shown as the dashed lines in panel a) and a priori HCN profiles (dashed lines in panel d). The retrieved profiles are shown as solid lines with color codes corresponding to the a priori. The HCN (resp. temperature) a priori is also shown in panels (b) (resp. (c)) as a grey dashed line.

These modelling variants yielded a suite of solution profiles, whose dispersion is a good indicator on the actual error bars on the retrieved temperature and HCN profile (Supplementary Figure 4). Beyond the well-known characteristics of Titan's middle atmosphere thermal structure such as the stratopause near 270 km, robust new features of all retrieved profiles are (i) the existence of a quasi-isothermal region over 450-600 km (already apparent in the CIRS-based a priori) and (ii) the existence of a ~140 K mesopause, located at 800±50 km altitude depending on the a priori. Error bars on the retrieved temperatures remain below 2-3 K up to 800 km; above this level they increase with the progressive lack of sensitivity of the measurements to ultimately reflect the a priori errors. The HCN profile (Fig. 3) is also very well defined (to within ~10-20 %) over 100-850 km, before error bars progressively increase. The robustness of the retrieved

HCN in previous ALMA observations to even factor >10 changes in the a priori was noted by [44] (their Fig. 2). The nominal HCN mixing profile we determine reaches a 6.3 x $10^{-4}$ mixing ratio at 1000 km, in agreement with results from both Cassini/INMS[41] and Cassini/UVIS[42], respectively above and below 950 km (see Fig. 3).

The retrieved thermal profile was then used to derive the other molecular profiles, again for the equatorial region. The DCN profile was taken as the retrieved HCN profile, multiplied by a constant DCN/HCN ratio of (2.5±0.2) x $10^{-4}$ [44], i.e. assuming that the DCN profile follows that of HCN at all altitudes. $CH_3CN$ and $CH_3CCH$ were inverted using a priori profiles respectively taken from ground-based data[39] (extended to $10^{-5}$ at 1000 km and above) and CIRS observations[45]. For these three species, best fits of the (west) limb equatorial emission are presented in Supplementary Figure 3. Error bars on $CH_3CN$ and $CH_3CCH$ reflect the combined effects of noise, temperature profile uncertainty and a priori errors. They were estimated by allowing for factor-of-2 variations in the a priori, and using also the warmest / coldest temperature profile when performing the inversion.

The HNC J=4-3 line exhibits a purely Doppler-broadened profile (Supplementary Figure 5), and as such contains limited information on the HNC vertical profile, except for the fact that this precludes HNC to be present at too deep atmospheric levels. Based on HNC J=6-5, Ref. [34] found that for constant mixing profiles above some level, the cutoff altitude cannot be lower than 300 km. We refined this conclusion by testing photochemical profiles[14] for HNC. As the spatial distribution of HNC J=4-3 shows only mild latitudinal variations, we averaged the data over a concentric ring (0.285"-0.475", i.e. 2000-3350 km from Titan center) to improve S/N. We found (Supplementary Figure 5, top right) that several of the HNC models from Ref. [14], in particular the nominal one, overestimate the amount of HNC in the lower stratosphere, producing unobserved spectral wings. In contrast, the model in which the production of HNC by Galactic Cosmic Ray impact is omitted is essentially in agreement with the observed spectrum, requiring only a modest rescaling by a factor 1.15. Nonetheless, this line shape-based approach does not constrain the distribution of HNC at altitudes above the pressure-broadening levels. The latter can be determined by inspecting the observed variation of the line-integrated intensities with radial distance. Specifically, we compared the latter with model calculations for constant HNC profiles above some cutoff altitudes, using the nominal retrieved temperature profile (blue profile in Supplementary Figure 1). Supplementary Figure 5 (top left) shows the result, with cutoff altitudes of 600, 800, 1000, 1200 km, considering together all the HNC data. Too low (resp. too high) cutoff altitudes lead the models to predict an emission maximum at too small (resp. too large) radial distances. Part of the dispersion of the observational points in this panel is due to the spatial variation of the HNC emission. Repeating the exercise, both by (i) generating synthetic datasets with similar dispersion as the observed and (ii) splitting the HNC measurements in angular bins of 30° across the disk, provides a robustness test of the method and indicates that the bulk of HNC is located above 870±40 km altitude, implying that HNC is mostly a thermospheric species. Furthermore, Supplementary Figure 5 (top left) shows that the photochemical model from Ref. [14] without GCR also provides the correct radial dependence for the HNC emission. As this profile has more physical basis than a piecewise distribution, we adopt it as nominal profile, allowing for a vertical shift of ±40 km in its altitude scale (see Fig. 3). Using extreme temperature profiles does not change the altitude distribution of HNC by more than 10 km.

The situation is similar for the $HC_3N$ J=39-38 emission at 354.697 GHz, which – except in the high-northern and southern latitude regions where it is strongly enhanced – does not show Lorentzian wings. We performed the same study as for HNC on the radial distance variation of the $HC_3N$ emission (using the slightly stronger J=38-37 line at 345.609 GHz observed along with CO (3-2)), restricting ourselves to points of the maps having a declination offset within [± 0.2”], i.e. [± 0.55 Titan's radius] from disk center, in order to focus on the low-latitude regions. Fitting models in which $HC_3N$ is assumed to be uniformly mixed above some cutoff altitudes and again refitting synthetic datasets with an appropriate noise level indicates a best fit cutoff altitude of 630±30 km (Supplementary Figure 5, bottom left), implying that $HC_3N$ too is dominantly present in Titan's upper atmosphere, albeit less so than HNC. This is not surprising given that $HC_3N$, as measured by CIRS[12-13] has also a measurable abundance in Titan's stratosphere and lower mesosphere, with mixing ratios increasing from $6\times10^{-11}$ – $1\times10^{-7}$ over 150-450 km. Initializing the $HC_3N$ profile at these values and a large abundance ($2\times10^{-4}$) at z > 800 km, we determined from inversion an equatorial $HC_3N$ profile fitting both the shape of the J=39-38 line at equatorial west limb and the radial variation of the line-integrated J=38-37 lines fluxes. In contrast (Supplementary Figure 5, bottom right), the nominal $HC_3N$ profile from Ref. [14] does not fit any either of the two observables, indicating that this profile is overabundant at lower altitudes and under-abundant in the upper atmosphere. Our $HC_3N$ solution profile has a 1000 km abundance as high as ~$5 \times 10^{-5}$. This agrees rather well with the global value reported from Cassini/INMS[46] ($3.2 \times 10^{-5}$ at 1025 km, after correction for possible effects of wall adsorption/desorption), and with the Cassini/UVIS $HC_3N$ abundance over 850-950 km [42]. We reiterate that our conclusions pertain to the equatorial $HC_3N$ profile only.

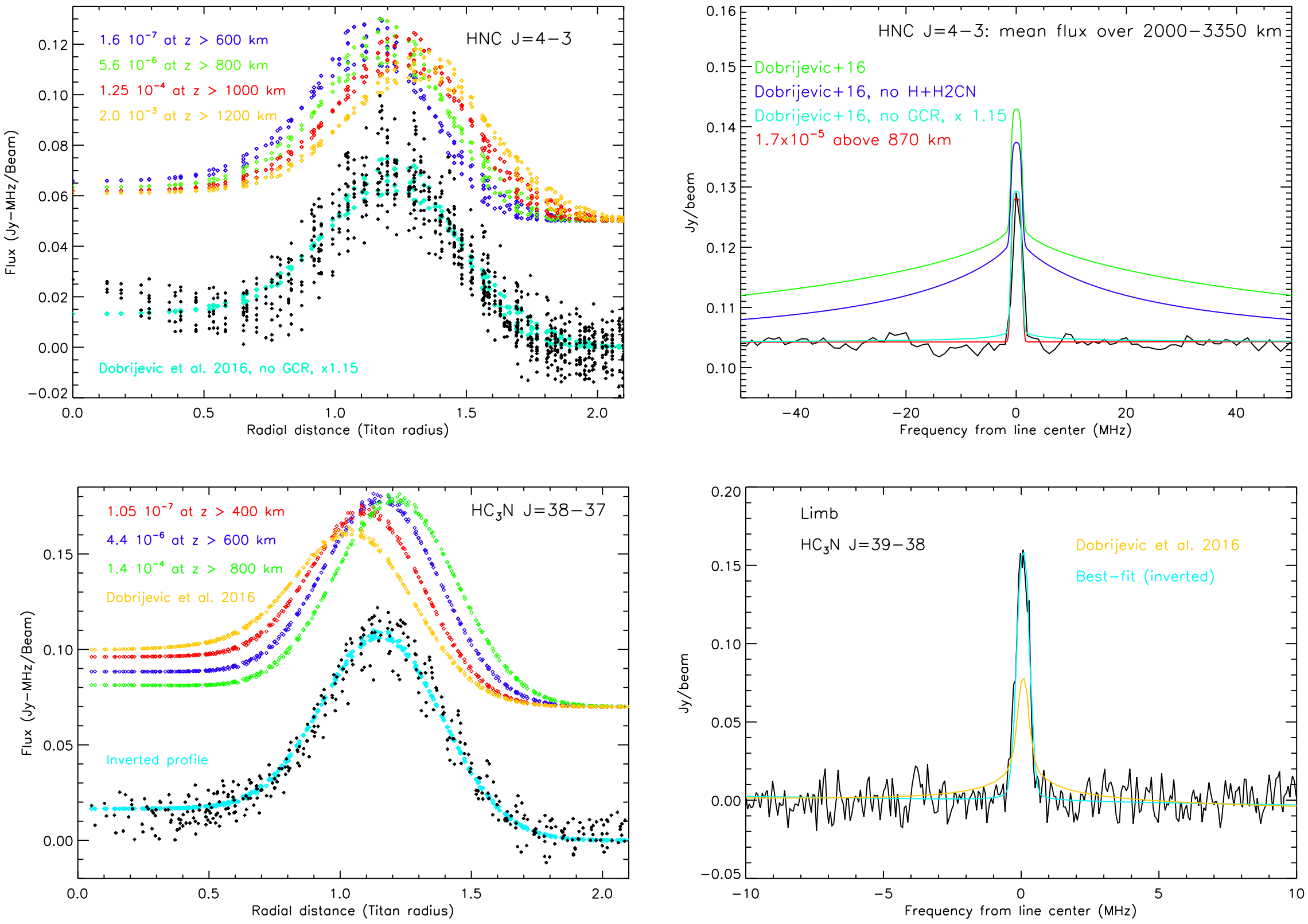

**Supplementary Figure 5. Determination of the vertical profile of HNC and $HC_3N$ from combination of line shape (right panels) and radial dependence (left panels) information. Left panels**: line-integrated fluxes (Jy-MHz per beam) for HNC (4-3) and $HC_3N$ (38-37) as a function of distance (in Titan radius) from Titan's center. For HNC, all points on the map are considered, while for $HC_3N$, only low-latitude regions are considered (see text). These radial profiles are compared both with piecewise profiles (HNC or $HC_3N$ assumed to be uniformly mixed above some cutoff altitudes) and with inverted ($HC_3N$) or physically-based profiles (models from Dobrijevic et al.[14] for HNC and $HC_3N$). **HNC:** For HNC, the piece-wise profiles have the following abundances: $1.6\times10^{-7}$ above 600 km, $5.6\times10^{-6}$ above 800 km, $1.25\times10^{-4}$ above 1000 km and $2.0\times10^{-3}$ above 1200 km. The models are shifted upwards by 0.05 units for clarity, and the "noise" in the models results from the non-circular beam. The best fit for piecewise models is achieved for $q_{HNC} = 1.7\times10^{-5}$ above 870 km. The HNC profile in which the dissociation by Galactic Cosmic Ray (GCR) is turned off (black profile in Fig. 26 of Ref. [14]), when multiplied by a modest factor of 1.15, also provides an excellent match to the radial dependence of the emission. **Top right panel**: HNC J = (39-38) line, extracted on an annulus ranging from 2000-3350 km from Titan center. The previous model with no GCR contribution matches the line shape; in contrast, other models from Ref. [14] (red and blue profiles in their Fig. 26) have too much HNC in the lower mesosphere below 600 km and produce unobserved line-wings. **$HC_3N$:** For $HC_3N$, the piecewise profiles have the following abundances: $1.05\times10^{-7}$ above 400 km, $4.4\times10^{-6}$ above 600 km, and $1.4\times10^{-4}$ above 800 km. The models are shifted upwards by 0.07 units for clarity. The best fit for piecewise models is achieved for $q_{HC3N} = 7\times10^{-6}$ above 630 km. Due to the selection criterion for the $HC_3N$ data, the associated profile pertains to the equatorial region only. **Bottom right panel.** $HC_3N$ J=39-38 line on the equatorial west limb, compared to best fit model (light blue) resulting from inversion of the line profile and including constraints from the radial dependence of the line-integrated $HC_3N$ (38-37) fluxes at low latitude. In contrast, the nominal $HC_3N$ profile of Ref. [14] (their Fig. 10, red line) fails at matching both observables (yellow symbols in the two bottom panels), indicating that that profile has too much $HC_3N$ at low altitudes and not enough at high altitudes.

Figure 3 summarizes the retrieved profiles for all six molecules used to probe the winds. Error bars are indicated in the range where information is available. For HNC and $HC_3N$, uncertainties are best expressed as a vertical shift of ±40 km and ±30 km in the profiles, for reasons explained above. These profiles are compared to characteristic profiles locally derived from Cassini/UVIS at 6°S (T41I, Ref. [42]), INMS[41,46], and CIRS. For the latter, we show HCN and $CH_3CCH$ profiles at 9°S from May 2016 (T119, Ref. [45]); for $HC_3N$, not clearly detected in the latter observations, we use instead a 5°N profile from Jan. 2007 (T23, Ref. [47]). Also shown in Fig. 3 are disk-averaged HCN and $HC_3N$ profiles derived from 2012-2015 ALMA observations ([33], their Fig. 10). Figure 3 highlights the overall agreement between these profiles, giving in particular great confidence in our HCN and $HC_3N$ profiles in the middle/upper atmosphere. A noteworthy difference is that "local" molecular profiles (from Cassini CIRS and UVIS) show small-scale vertical variations (see e.g. Figs. 3-4 of Ref. [47], and Fig. 26 of Ref. [42]), not predicted in photochemical models[14] and presumably of dynamical origin; those are particularly visible in Fig. 3 for the HCN and $HC_3N$ profiles derived from UVIS.

## 4) Wind structure and wind models

In Supplementary Figure 6, the line-of-sight wind velocities are plotted as a function of the "algebraic" projected distance to Titan's axis – by definition, the latter is positive (resp. negative) sky-east (resp. west) of the polar axis, and the distance is expressed in units of Titan's "effective" radius, i.e. the (species-dependent) radial distance at the altitude of the wind contribution function (WCF)-weighted mean probed altitude. (Given the strong variation of the HCN WCF between nadir and limb, HCN LOS winds within the disk are not included). In this type of representation (and ignoring for now beam convolution effects), a solid-body rotation appears as a straight line with slope equal to $v_{eq}$ cos(β), where β = 26°N is the sub-Earth latitude. (In contrast, such plots do not permit one to investigate latitudinal dependences). Best fit and error bars for $v_{eq}$ were obtained from error-bar weighted fits of the individual measurements. In Supplementary Figure 6**,** LOS winds from different molecules are compared two-by-two, and linear fits are over-plotted with best fit $v_{eq}$ values indicated. This figure highlights several facts: (i) wind values from $CH_3CN$, $CH_3CCH$, and DCN show excellent consistency (despite the larger uncertainties on the latter two species), both in their structure and in the $v_{eq}$ value (191-206 m/s) (ii) HNC winds are clearly different in speed and structure; although a solid-body rotation model leads to a best fit $v_{eq}$ = 297 m/s, they do not align well on a straight line, instead showing a deviation from the solid body rotation to higher velocities at large distances, i.e. an equatorial jet (iii) The $v_{eq}$ values for the HCN main line and $HC_3N$ winds are virtually identical (287-290 m/s), and the structures seen in their maps seem intermediate between those of $CH_3CN$ and HNC (although HCN exhibits a stronger "tongue" at large distances than $HC_3N$) (iv) although the HCN main line nominally probes the atmosphere

~200 km higher than the satellite lines (see Fig. 3), the associated $v_{eq}$ values are only marginally different. The last panel of Supplementary Figure 6 shows the variation of the fitted equatorial velocity with altitude, as indicated by the WCF for each molecule. These findings are consistent with the three altitude regions outlined above and with a general increase of winds with altitude, with a marked gradient over 350-600 km. Still, the vertical shear du/dz (using $u = v_{eq}$) is small enough that the Richardson number $Ri = N^2 / (du/dz)^2$, where $N$ is the Brunt–Väisälä frequency, rises from ~20 at 300 km to ~1000 at 1000 km. This is safely above the threshold Ri < ¼ for turbulence.

The above description is approximate, especially because it ignores beam-convolution effects. In addition, the data indicate more complex structure than solid-body rotation, with a progressive onset of the jet regime above ~600 km. To estimate the wind regime indicated by the data, we simulated the expected LOS Doppler shifts, testing various cases for the latitudinal distribution of winds. The expected Doppler shifts were calculated locally by considering a globe with radius defined by the mean probed altitude and projecting the zonal winds accordingly, accounting for the viewing geometry with sub-earth latitude at 26°N. In doing so, we also included the minor effect of Titan's solid body rotation with orbital period of 15.95 days (giving e.g. a 16 m/s equatorial speed at 990 km altitude, the nominal altitude probed by HNC). Only purely zonal winds were considered, e.g. no meridional components were included. The modelled local LOS Doppler shifts were then weighted by the local line intensity – i.e. the line emission signal integrated over the spectral interval used to measure the winds in the data – and finally convolved by the instrumental beam. We performed this exercise for winds measured in $CH_3CN$, HNC and $HC_3N$, noting that this is still an approximate way of modelling the data. Indeed, given the vertical extent of the wind contribution functions, especially for $HC_3N$ and HCN, the only completely rigorous approach would consist of testing 2D (latitude, altitude) wind fields and generate the associated synthetic line profiles, incorporating the altitude-dependent Doppler shifts in the absorption coefficients before performing the radiative transfer. Such synthetic profiles would finally be refit for Doppler shifts in the same manner as the actual data. This complex task, particularly needed for $HC_3N$ and HCN, is left for the future. With this caveat in mind, Supplementary Figure 7 shows synthetic wind models for $CH_3CN$, HNC and $HC_3N$ that provide a satisfactory match to the data shown in Fig. 2. As is clear from the first panel of Supplementary Figure 7, the zonal wind magnitude and its latitudinal dependence are different between the three species, indicative of an evolution of the wind regime with altitude.

Meridional winds are not obviously present in the data. Assuming equator-to-pole or pole-to-pole circulation, a meridional circulation would be best visible at mid-latitudes along the central meridian. Only the $CH_3CN$ data are suited for that, and based on the wind contours in Fig. 2, we estimate an upper limit of ~25 m/s for such winds. Similarly, day-to-night flows would be best captured from Doppler shifts in the polar region (e.g. receding winds at the poles for SSAS winds). Our data do not give any consistent evidence for this either, with an upper limit estimated to ~30 m/s.

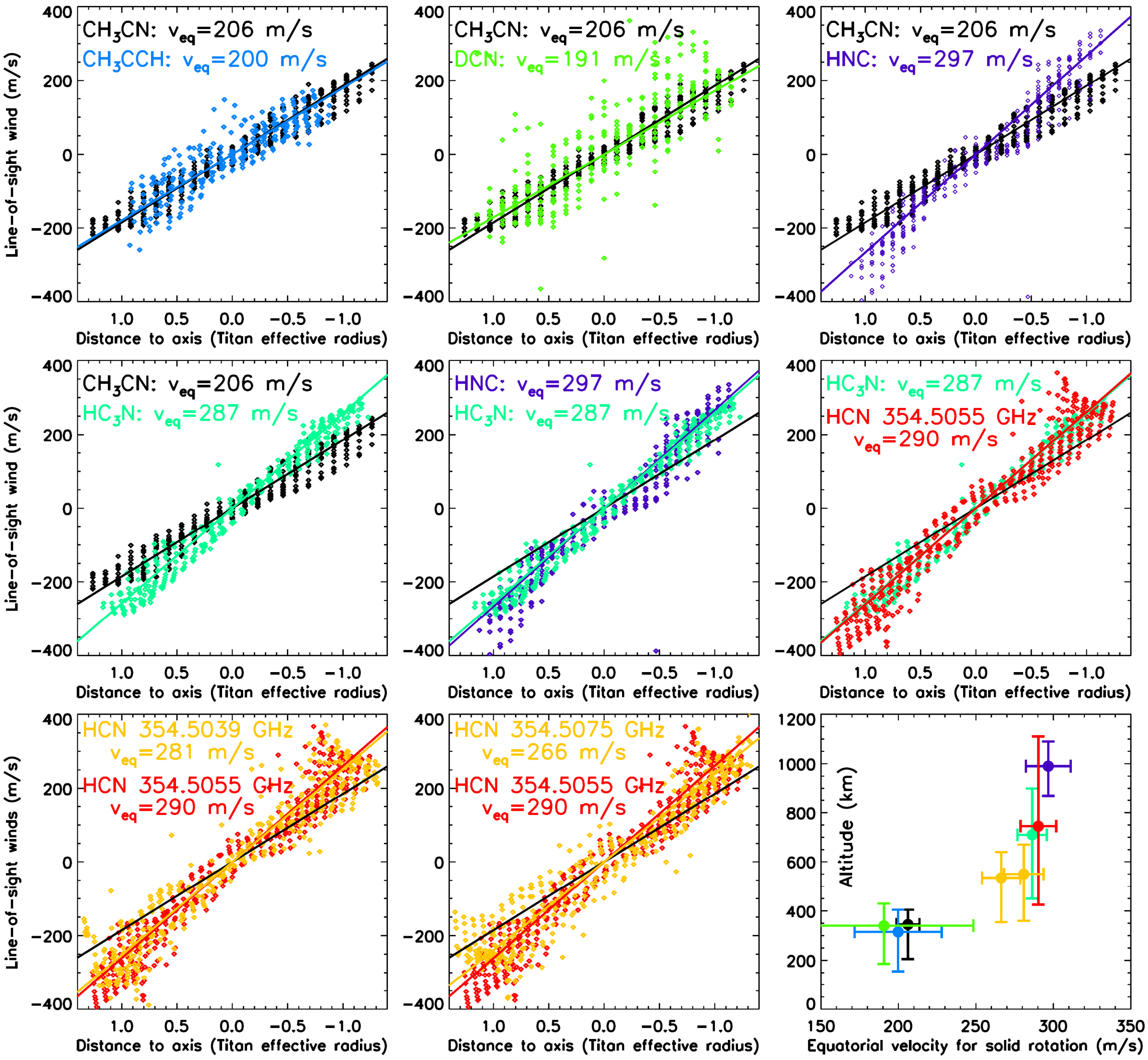

**Supplementary Figure 6. Line-of-sight wind velocities as a function of projected distance from polar axis**. The distance is expressed in units of "Titan effective radius", defined as the radial distance at the WCF-weighted mean probed altitude (e.g. 3565 km for HNC). By convention, the distance is positive (resp. negative) for points located sky-east (resp. west) from the polar axis. Straight lines show simple linear fits to the data assuming solid-body rotation (see text) with the corresponding equatorial velocities indicated. The fit for $CH_3CN$ (black line) is repeated on all panels for comparison. The last panel shows the variation of the fitted equatorial velocity with altitude. In that panel, the color code is used consistently with the other panels, and the vertical error bars indicate the extent of the wind contribution functions (see Fig. 3).

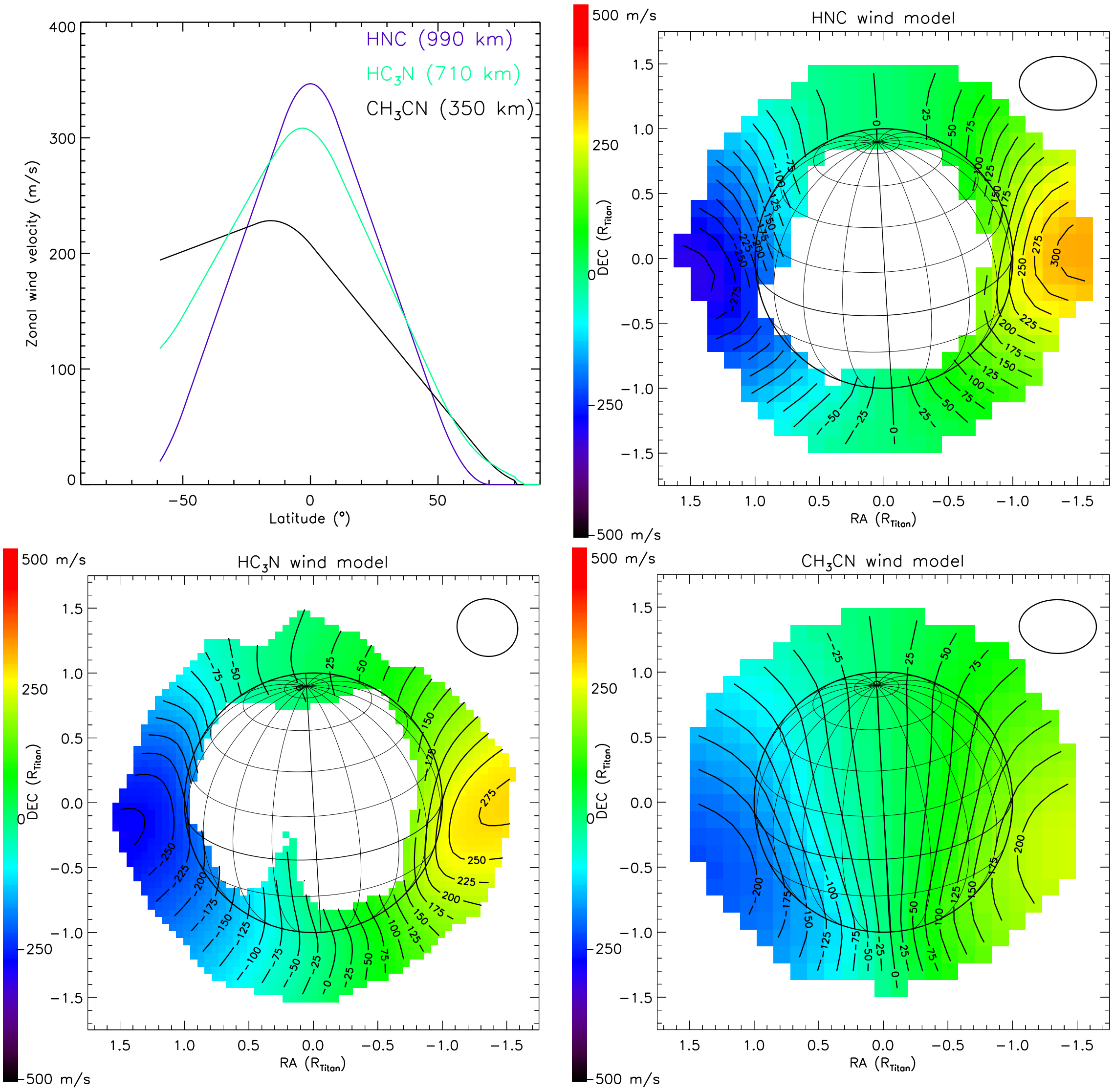

**Supplementary Figure 7. Modelled wind structure.** Simulated Doppler shifts for HNC, $HC_3N$ and $CH_3CN$, using the respective latitude profiles of the zonal wind shown in the first panel. Local line-of-sight winds are weighted by line intensities and convolved by the instrumental beam. Titan's solid-body rotation is included. These figures approximate the observed maps for these species (see Fig. 2).